\documentclass{revtex4}
\usepackage{amssymb}
\usepackage{amsmath}

\input{tcilatex}

\begin{document}

\title[ ]{Constant curvature f(R) gravity minimally coupled with Yang-Mills
field }
\author{S. Habib Mazharimousavi}
\email{habib.mazhari@emu.edu.tr}
\author{M. Halilsoy}
\email{mustafa.halilsoy@emu.edu.tr}
\author{T. Tahamtan}
\email{tayabeh.tahamtan@emu.edu.tr}
\affiliation{Department of Physics, Eastern Mediterranean University, G. Magusa, north
Cyprus, Mersin 10, Turkey. }

\begin{abstract}
We consider the particular class of $f(R)$ gravities minimally coupled with
Yang - Mills (YM) field in which the Ricci scalar $=R_{0}=$ constant in all
dimensions $d\geq 4$. Even in this restricted class the spacetime has
unlimited scopes determined by an equation of state of the form $%
P_{eff}=\omega \rho $. Depending on the distance from the origin (or horizon
of a black hole) the state function $\omega \left( r\right) $ takes
different values. It is observed that $\omega \rightarrow \frac{1}{3}$ (the
ultra relativistic case in $4$\ - dimensions) and $\omega \rightarrow -1$
(the cosmological constant) are the limiting values of our state function $%
\omega \left( r\right) $ in a spacetime centered by a black hole. This
suggests that having a constant $\omega $ throughout spacetime around a
charged black hole in $f(R)$ gravity with constant scalar curvature is a
myth.
\end{abstract}

\maketitle

\section{INTRODUCTION}

For a number of reasons, ranging from dark energy and accelerated expansion
of the universe to astronomical tests, modified version of general
relativity gained considerable interest in recent times.  $f(R)$ gravity, in
particular, attracted much attention in this context (see \cite{1} for
comprehensive reviews of the subject). The reason for this trend may be
attributed to the dependence of its Lagrangian on the Ricci scalar alone, so
that it can be handled relatively simpler in comparison with the higher
order curvature invariants. Depending on the structure of the function $f(R)$
the nonlinearity creates curvature sources which may be interpreted as
'sources without sources', manifesting themselves in the Einstein equations.
Beside these curvature (or geometrical) sources there may be true physical
sources that contribute together with the former to determine the total
source in the problem. It should be added that owing to highly nonlinear
structure of the underlying field equations attaining exact solutions is not
an easy task at all. In spite of all odds many exact solutions have been
obtained from ab initio assumed $f(R)$ functions. To recall an example we
refer to the choice $f(R)=R^{N}$, ($N=$ an arbitrary number) which attains
an electromagnetic - like curvature source, so that $N\neq 1$ can be
interpreted as an 'electric charge without charge' \cite{2}. That is, the
resulting geometry becomes equivalent to the Reissner-Nordstrom (RN)
geometry in a spherically symmetry metric ansatz of Einstein's gravity. This
particular example reveals that the failure of certain tests related to
Solar System / Cosmology in $f(R)$ gravity is accountable by the curvature
sources in the Einstein Hilbert action. Equivalence with $f(R)=R+$(scalar
fields) provides another such example beside the electromagnetic one. More
recently we obtained a large class of non-analytical $f(R)$ gravity
solutions minimally coupled with Yang-Mills (YM) field \cite{3}. Even more
to this the YM field was allowed to be a nonlinear theory in which the
power-YM constitutes a particular example in all higher dimensions. In
particular, in $d=6$, $\ f(R)=\sqrt{R}$ solves the Einstein-Yang-Mills (EYM)
system exactly. For $d=4$ \ our solution for nonabelian gauge reduces to an
abelian one which may be considered as an Einstein-Maxwell (EM) solution 
\cite{4}.

Previously $f(R)$ gravity coupled non-minimally with Yang-Mills and Maxwell
matter sources have been considered \cite{5}. In this paper we consider a
particular class within minimally coupled YM field in $f(R)$ gravity with
the conditions that the scalar curvature $R=R_{0}=$ constant and the trace
of the YM energy-momentum tensor is zero. (To see other black hole solutions
with matter in $f(R)$ gravity we refer to Ref. \cite{6}). Contrary to our
expectations this turns out to be a non-trivial class with far-reaching
consequences. Our spacetime is chosen spherically symmetric to be in accord
with the spherically symmetric Wu-Yang ansatz for the YM field. The field
equations admit exact solutions in all dimensions $d\geq 4$ with the
physical parameters; mass ($m$) of the black hole, YM charge ($Q$) and the
scalar curvature ($R_{0}$) of the space time. In this picture we note that
cosmological constant arises automatically as proportional to $R_{0}$. From
physics stand point, considering equation of state in the form $%
P_{eff}=\omega \rho $, with effective pressure ($P_{eff}$) and density ($%
\rho $), important results are obtained as follow: For a critical value of $%
r=r_{c}$ we have $-1<\omega \left( r\right) <0$ for $r>r_{c}$ and $0<\omega
\left( r\right) <\frac{1}{3}$ for $r<r_{c}$. Remarkably this amounts to a
sign shift in the effective pressure to account for the accelerated
expansion of a universe centered by a charged black hole. In general the
critical distance is thermodynamically unstable so that the universe
undergoes the phase of accelerated expansion beyond that particular
distance. Absence of the phantom era (i.e. $\omega <-1$) is also manifest.
Alternatively, in the limit $r\rightarrow 0$ it yields $\omega \rightarrow 
\frac{1}{3}$ (i.e. $4-$dimensional ultra relativistic case), while for $%
r\rightarrow \infty $ we have $\omega \rightarrow -1$, the case of a pure
cosmological constant. Let us note that the latter case corresponds to
vanishing of the YM field. Stated otherwise, in the overall space time we do
not have a fixed value for $\omega $. Depending on the distance from the
center (or horizon) of a black hole we have a varying state parameter $%
\omega \left( r\right) $. The same argument in the
Friedmann-Robertson-Walker (FRW) version of the theory implies that beyond a
critical time $t=t_{c}$, $\omega \left( t\right) $ changes its role and a
different type of matter becomes active. It is known that for the normal and
dark matters which provide clustering both the weak energy condition (WEC)
and the strong energy condition (SEC) must be satisfied. In the case of dark
energy on the other hand WEC is satisfied while SEC is violated. In Appendix
A we analyze the energy conditions thoroughly covering all dimensions.
Although our metric ansatz is chosen to be spherically symmetric so that the
constant scalar curvature $R_{0}>0$, in order to prepare ground for the
topological black holes we consider the case of $R_{0}<0$ as well.

Organization of the paper is as follows. In Sec. II we introduce our
formalism and present exact solutions. The analysis of our solution with
thermodynamical functions is considered in Sec. III. We complete the paper
with Conclusion which appears in Sec. IV.

\section{The formalism and solution for $R=$constant.}

We choose the action as (Our unit convention is chosen such that $c=G=1$ so
that $\kappa =8\pi $)%
\begin{equation}
S=\int d^{d}x\sqrt{-g}\left[ \frac{f\left( R\right) }{2\kappa }+\mathcal{L}%
\left( F\right) \right]
\end{equation}%
in which $f\left( R\right) $ is a real function of Ricci scalar $R$ and $%
L\left( F\right) $ is the nonlinear YM Lagrangian with $F=\frac{1}{4}%
tr\left( F_{\mu \nu }^{\left( a\right) }F^{\left( a\right) \mu \nu }\right)
. $ Obviously the particular choice $\mathcal{L}\left( F\right) =-\frac{1}{%
4\pi }F$ will reduce to the case of standard YM theory. The YM field $2-$%
form components are given by 
\begin{equation}
\mathbf{F}^{\left( a\right) }=\frac{1}{2}F_{\mu \nu }^{\left( a\right)
}dx^{\mu }\wedge dx^{\nu }
\end{equation}%
with the internal index $(a)$ running over the degrees of freedom of the
nonabelian YM gauge field. Variation of the action with respect to the
metric $g_{\mu \nu }$ gives the EYM field equations as%
\begin{equation}
f_{R}R_{\mu }^{\nu }+\left( \square f_{R}-\frac{1}{2}f\right) \delta _{\mu
}^{\nu }-\nabla ^{\nu }\nabla _{\mu }f_{R}=\kappa T_{\mu }^{\nu }
\end{equation}%
in which 
\begin{gather}
T_{\mu }^{\nu }=\mathcal{L}\left( F\right) \delta _{\mu }^{\nu }-tr\left(
F_{\mu \alpha }^{\left( a\right) }F^{\left( a\right) \nu \alpha }\right) 
\mathcal{L}_{F}\left( F\right) , \\
\mathcal{L}_{F}\left( F\right) =\frac{d\mathcal{L}\left( F\right) }{dF}. 
\notag
\end{gather}%
Our notation here is as follows: $f_{R}=\frac{df\left( R\right) }{dR}$, $%
\square f_{R}=\nabla _{\mu }\nabla ^{\mu }f_{R}=\frac{1}{\sqrt{-g}}\partial
_{\mu }\left( \sqrt{-g}\partial ^{\mu }\right) f_{R}$ , $R_{\mu }^{\nu }$ is
the Ricci tensor and 
\begin{equation}
\nabla ^{\nu }\nabla _{\mu }f_{R}=g^{\alpha \nu }\left( f_{R}\right) _{,\mu
;\alpha }=g^{\alpha \nu }\left[ \left( f_{R}\right) _{,\mu ,\alpha }-\Gamma
_{\mu \alpha }^{m}\left( f_{R}\right) _{,m}\right] .
\end{equation}%
\ The trace of the EYM equation (3) yields%
\begin{equation}
f_{R}R+\left( d-1\right) \square f_{R}-\frac{d}{2}f=\kappa T
\end{equation}%
in which $T=T_{\mu }^{\mu }.$ The $SO\left( d-1\right) $ gauge group YM
potentials are given by 
\begin{align}
\mathbf{A}^{(a)}& =\frac{Q}{r^{2}}C_{\left( i\right) \left( j\right)
}^{\left( a\right) }\ x^{i}dx^{j},\text{ \ \ }Q=\text{YM magnetic charge, \ }%
r^{2}=\overset{d-1}{\underset{i=1}{\sum }}x_{i}^{2}, \\
2& \leq j+1\leq i\leq d-1,\text{ \ and \ }1\leq a\leq \left( d-2\right)
\left( d-1\right) /2,  \notag \\
x_{1}& =r\cos \theta _{d-3}\sin \theta _{d-4}...\sin \theta _{1},\text{ }%
x_{2}=r\sin \theta _{d-3}\sin \theta _{d-4}...\sin \theta _{1},  \notag \\
\text{ }x_{3}& =r\cos \theta _{d-4}\sin \theta _{d-5}...\sin \theta _{1},%
\text{ }x_{4}=r\sin \theta _{d-4}\sin \theta _{d-5}...\sin \theta _{1}, 
\notag \\
& ...  \notag \\
x_{d-2}& =r\cos \theta _{1},  \notag
\end{align}%
in which $C_{\left( b\right) \left( c\right) }^{\left( a\right) }$ are the
non-zero structure constants of $\frac{\left( d-1\right) \left( d-2\right) }{%
2}-$parameter Lie group $\mathcal{G}$ \cite{7,8}. The metric ansatz is
spherically symmetric which reads%
\begin{equation}
ds^{2}=-A\left( r\right) dt^{2}+\frac{dr^{2}}{A\left( r\right) }%
+r^{2}d\Omega _{d-2}^{2},
\end{equation}%
with the only unknown function $A\left( r\right) $ and the solid angle
element 
\begin{equation}
d\Omega _{d-2}^{2}=d\theta _{1}^{2}+\underset{i=2}{\overset{d-2}{\tsum }}%
\underset{j=1}{\overset{i-1}{\tprod }}\sin ^{2}\theta _{j}\;d\theta _{i}^{2},
\end{equation}%
with%
\begin{equation*}
0\leq \theta _{d-2}\leq 2\pi ,0\leq \theta _{i}\leq \pi ,\text{ \ \ }1\leq
i\leq d-3.
\end{equation*}%
Variation of the action with respect to $\mathbf{A}^{\left( a\right) }$
implies the YM equations 
\begin{equation}
\mathbf{d}\left[ ^{\star }\mathbf{F}^{\left( a\right) }L_{F}\left( F\right) %
\right] +\frac{1}{\sigma }C_{\left( b\right) \left( c\right) }^{\left(
a\right) }L_{F}\left( F\right) \mathbf{A}^{\left( b\right) }\wedge ^{\star }%
\mathbf{F}^{\left( c\right) }=0,
\end{equation}%
in which $\sigma $ is a coupling constant and $^{\star }$ means duality. One
may show that the YM invariant satisfies 
\begin{equation}
F=\frac{1}{4}tr\left( F_{\mu \nu }^{\left( a\right) }F^{\left( a\right) \mu
\nu }\right) =\frac{\left( d-2\right) \left( d-3\right) Q^{2}}{4r^{4}}
\end{equation}%
and%
\begin{equation}
tr\left( F_{t\alpha }^{\left( a\right) }F^{\left( a\right) t\alpha }\right)
=tr\left( F_{r\alpha }^{\left( a\right) }F^{\left( a\right) r\alpha }\right)
=0,
\end{equation}%
while%
\begin{equation}
tr\left( F_{\theta _{i}\alpha }^{\left( a\right) }F^{\left( a\right) \theta
_{i}\alpha }\right) =\frac{\left( d-3\right) Q^{2}}{r^{4}},
\end{equation}%
which leads us to the exact form of the energy momentum tensor%
\begin{equation}
T_{\mu }^{\nu }=\text{diag}\left[ \mathcal{L},\mathcal{L},\mathcal{L}-\frac{%
\left( d-3\right) Q^{2}}{r^{4}}\mathcal{L}_{F},\mathcal{L}-\frac{\left(
d-3\right) Q^{2}}{r^{4}}\mathcal{L}_{F},...,\mathcal{L}-\frac{\left(
d-3\right) Q^{2}}{r^{4}}\mathcal{L}_{F}\right] .
\end{equation}%
Here the trace of $T_{\mu }^{\nu }$ becomes%
\begin{equation}
T=T_{\mu }^{\mu }=d\mathcal{L}-4F\mathcal{L}_{F},
\end{equation}%
and therefore with Eq. (3) we find%
\begin{equation}
f=\frac{2}{d}\left[ f_{R}R+\left( d-1\right) \square f_{R}-\kappa \left( d%
\mathcal{L}-4F\mathcal{L}_{F}\right) \right] .
\end{equation}%
To proceed further we set the trace of energy momentum tensor to be zero
i.e., 
\begin{equation}
d\mathcal{L}-4F\mathcal{L}_{F}=0
\end{equation}%
which leads to a power Maxwell Lagrangian \cite{9} 
\begin{equation}
\mathcal{L}=-\frac{1}{4\pi }F^{\frac{d}{4}}.
\end{equation}%
Here for our convenience the integration constant is set to be $-\frac{1}{%
4\pi }.$ On the other hand, the constant curvature $R=R_{0}$, and the zero
trace condition together imply%
\begin{equation}
f^{\prime }\left( R_{0}\right) R_{0}-\frac{d}{2}f\left( R_{0}\right) =0.
\end{equation}%
This equation admits%
\begin{equation}
f\left( R_{0}\right) =R_{0}^{\frac{d}{2}},
\end{equation}%
where the integration constant is set to be one. One can easily write the
Einstein equations as 
\begin{equation}
G_{\mu }^{\nu }=\kappa \text{\ }\tilde{T}_{\mu }^{\nu }
\end{equation}%
where 
\begin{eqnarray}
\text{ \ \ }\tilde{T}_{\mu }^{\nu } &=&\frac{2R_{0}}{f\left( R_{0}\right) d}%
T_{\mu }^{\nu }-\frac{\Lambda _{eff}}{\kappa }\delta _{\mu }^{\nu },\text{ }
\\
\text{ }\Lambda _{eff} &=&\frac{\left( d-2\right) R_{0}}{2d},
\end{eqnarray}%
and in which $T_{\mu }^{\nu }$ is given by (4). The constancy of the Ricci
scalar amounts to 
\begin{equation}
-\frac{r^{2}A^{\prime \prime }+2\left( d-2\right) rA^{\prime }+\left(
d-2\right) \left( d-3\right) \left( A-1\right) }{r^{2}}=R_{0}
\end{equation}%
which yields%
\begin{equation}
A=1-\frac{R_{0}}{d\left( d-1\right) }r^{2}-\frac{m}{r^{d-3}}+\frac{\sigma }{%
r^{d-2}},
\end{equation}%
where $\sigma $ and $m$ are two integration constants. From the Einstein
equations one identifies the constant $\sigma $\ as%
\begin{equation}
\sigma =\frac{8}{d\left( d-2\right) R_{0}^{\frac{d-2}{2}}}\left( \frac{%
\left( d-3\right) \left( d-2\right) Q^{2}}{4}\right) ^{\frac{d}{4}}.
\end{equation}%
In the next section we investigate physical properties of our solution in
all dimensions.

\section{Analysis of the solution}

\subsection{$4-$dimensions}

\subsubsection{Thermodynamics}

In $4-$dimensions, we know that the nonabelian $SO(3)$ gauge field coincides
with the abelian $U(1)$ Maxwell field \cite{4}. Due to its importance we
shall study the $4-$dimensional case separately and give the results
explicitly. First of all, in $4-$dimensions the metric function becomes 
\begin{equation}
A=1-\frac{R_{0}}{12}r^{2}-\frac{m}{r}+\frac{Q^{2}}{2R_{0}r^{2}},\text{ \ \ \
\ \ \ }0<\left\vert R_{0}\right\vert <\infty 
\end{equation}%
and the form of action reads as%
\begin{equation}
S=\int d^{4}x\sqrt{-g}\left[ \frac{f\left( R\right) }{2\kappa }+\mathcal{L}%
\left( F\right) \right] 
\end{equation}%
in which 
\begin{equation}
f\left( R\right) =R^{2},
\end{equation}%
with $R=R_{0}$ and 
\begin{equation}
\mathcal{L}\left( F\right) =-\frac{1}{4\pi }F.
\end{equation}%
By assumption, $R_{0}$ gets positive / negative values and the resulting
spacetime becomes de-Sitter / anti de-Sitter, type in $f(R)=R^{2}$ theory
respectively, with effective cosmological constant $\Lambda _{eff}=\frac{%
R_{0}}{4}$. Let us add that in order to preserve the sign of the charge term
in (27) we must abide by the choice $R_{0}>0$. However, simultaneous limits $%
Q^{2}\rightarrow 0$ and $R_{0}\rightarrow 0$, so that $\frac{Q^{2}}{R_{0}}%
=\lambda _{0}=$constant, leads also to an acceptable solution within $%
f\left( R\right) $ gravity \cite{2}. It is not difficult to see here that $m$
is the ADM mass of the resulting black hole. Viability of the pure $%
f(R)=R^{2}$ model which has recently been considered critically \cite{10} is
known to avoid the Dolgov-Kawasaki instability \cite{11}. Further, in the
late time behaviour of the expanding universe (i.e. for $r\rightarrow \infty 
$) it asymptotes to the de Sitter / anti de Sitter form. With reference to 
\cite{10} we admit that sourceless $f(R)=R^{2}$ model doesn't possess a good
record as far as the Solar System tests are concerned. Herein we have
sources and wish to address the universe at large. Now, we follow \cite{12}
to give the form of the entropy akin to the possible black hole solution.
From the area relation the entropy of the modified gravity with constant
curvature is given by%
\begin{equation}
S=\frac{\mathcal{A}_{h}}{4G}f^{\prime }(R_{0})
\end{equation}%
which upon insertion from (19) becomes%
\begin{equation}
S=\frac{\mathcal{A}_{h}}{2GR_{0}}f\left( R_{0}\right) =2\pi R_{0}r_{h}^{2}
\end{equation}%
where $r_{h}$ indicates the event horizon. The Hawking temperature and heat
capacity are given respectively by 
\begin{equation}
T_{H}=\frac{A^{\prime }(r_{h})}{4\pi }=\frac{%
4R_{0}r_{h}^{2}-2Q^{2}-R_{0}^{2}r_{h}^{4}}{16\pi R_{0}r_{h}^{3}},
\end{equation}%
and$\bigskip $%
\begin{equation}
C_{Q}=T_{H}\frac{\partial S}{\partial T_{H}}=\frac{4\pi R_{0}r_{h}^{2}\left(
R_{0}^{2}r_{h}^{4}-4R_{0}r_{h}^{2}+2Q^{2}\right) }{\left(
R_{0}^{2}r_{h}^{4}+4R_{0}r_{h}^{2}-6Q^{2}\right) }.
\end{equation}%
Here we note that for the case of zero YM charge ($Q=0$) one finds 
\begin{equation}
C_{Q}=T_{H}\frac{\partial S}{\partial T_{H}}=4\pi R_{0}r_{h}^{2}\frac{\left(
R_{0}r_{h}^{2}-4\right) }{\left( 4+R_{0}r_{h}^{2}\right) }
\end{equation}%
which clearly shows from (34) that for $R_{0}>0,$ the YM source brings in
the possibility of having a phase change. This is depicted in Fig. 1. In
Fig.s 1A and 1C we plot the horizon radius versus mass for $Q=0$ and $Q=1$.
Similarly in Fig.s 1B and 1D we plot the heat capacity $C$ for $Q=0$ and $%
C_{Q}$ for $Q=1$ to see the drastic difference. It is observed that for $Q=0$
(Fig. 1B) the heat capacity is regular whereas for $Q=1$ (Fig. 1D), $C_{Q}$
is a discontinuous function signalling a phase change.

\subsubsection{Energy Conditions}

From the energy conditions (see Appendix, A2) the density and principal
pressures are given as

\begin{equation}
\rho =-\tilde{T}_{0}^{0}=\frac{1}{8\pi R_{0}}\left( F+\frac{1}{4}%
R_{0}^{2}\right) ,\text{ \ }p_{1}=\tilde{T}_{1}^{1}=-\frac{1}{8\pi R_{0}}%
\left( F+\frac{1}{4}R_{0}^{2}\right) ,\text{ \ }p_{i}=\tilde{T}_{i}^{i}=%
\frac{1}{8\pi R_{0}}\left( F-\frac{1}{4}R_{0}^{2}\right) ,\text{ \ \ \ }%
i=2,3.  \notag
\end{equation}%
These conditions imply that for $R_{0}\geq 0$, both the WEC and SEC are
satisfied. DEC implies, on the other hand, from (A7) that

\begin{equation}
P_{eff}=\frac{1}{3}\underset{i=1}{\overset{3}{\tsum }}\tilde{T}_{i}^{i}=%
\frac{1}{24\pi R_{0}}\left( F-\frac{3}{4}R_{0}^{2}\right) \geq 0,
\end{equation}%
which yields%
\begin{equation}
R_{0}\geq 0\text{ \ and \ }F\geq \frac{3}{4}R_{0}^{2}\rightarrow r\leq \sqrt[%
4]{\frac{2Q^{2}}{3R_{0}^{2}}}.
\end{equation}%
In addition to the energy conditions one can impose the causality condition
(CC) from (A9)

\begin{equation}
0\leq \frac{P_{eff}}{\rho }=\frac{\left( F-\frac{3}{4}R_{0}^{2}\right) }{%
3\left( F+\frac{1}{4}R_{0}^{2}\right) }<1,
\end{equation}%
which is satisfied if $F\geq \frac{3}{4}R_{0}^{2}$ or $r\leq \sqrt[4]{\frac{%
2Q^{2}}{3R_{0}^{2}}}.$

Finally, if one introduces a new parameter (as the equation of state
function $\omega $) by $\omega =\frac{P_{eff}}{\rho }$, one observes that in
the range for $0<r<\infty $ we have

\begin{equation}
-1\leq \omega <\frac{1}{3}.
\end{equation}%
In terms of the physical parameters, if 
\begin{equation}
\sqrt[4]{\frac{2Q^{2}}{3R_{0}^{2}}}\leq r
\end{equation}%
then $-1\leq \omega \leq 0$, and if 
\begin{equation}
\sqrt[4]{\frac{2Q^{2}}{3R_{0}^{2}}}>r
\end{equation}%
we have $0<\omega <\frac{1}{3}.$ It is clearly seen that the foregoing
bounds serve to define possible critical distances where the sign of the
effective pressure changes sign. This may be interpreted as changing phase
for example, from contraction to expansion or vice versa in a universe
centered by a black hole. We note that scaling the mass and distance by $%
R_{0}$ the results will not be affected. For this reason we set $R_{0}=1.$
From Eq. (34) we plot in Fig.2, $C_{Q},$ (with $R_{0}=1$) versus $r_{h}$ and 
$Q$. The shaded region \ for $r_{h}<r_{c}$ and $C_{Q}>0,$which lies below
the curve $r_{h}=r_{c}$ is the stable region outside the black hole. All the
rest with $C_{Q}<0$ is thermodynamically unstable region. Fig. 2 reveals
that except for a very narrow band of stability islands there is a vast
region of instability for $r_{c}$ at which the effective pressure turns sign
and continues into opposite pressure, i.e. expansion reverses into
contraction or vice versa.

\subsection{$d-$ dimensions}

\subsubsection{Thermodynamics }

In higher dimensions one obtains for the entropy and Hawking temperature the
following expressions (for $n\geq 2$) 
\begin{equation}
S=\left\{ 
\begin{array}{cc}
\frac{\left( 2n-1\right) n\pi ^{\frac{2n-1}{2}}r_{h}^{2\left( n-1\right)
}R_{0}^{n-1}}{4\Gamma \left( n+\frac{1}{2}\right) }, & d=2n \\ 
\frac{\left( 2n+1\right) n\pi ^{n}r_{h}^{2n-1}R_{0}^{\left( 2n-1\right) /2}}{%
4\Gamma \left( n+1\right) }, & d=2n+1%
\end{array}%
\right. ,
\end{equation}%
\begin{equation}
T_{H}=\left\{ 
\begin{array}{ll}
\frac{-1}{8n\pi r_{h}^{2n-1}R_{0}^{n-1}}\left[ \frac{4Q^{n}}{\left(
n-1\right) }\left( \frac{\left( 2n-3\right) \left( n-1\right) }{2}\right) ^{%
\frac{n}{2}}+\left( 6n-4n^{2}+R_{0}r_{h}^{2}\right) r_{h}^{2\left(
n-1\right) }R_{0}^{n-1}\right] , & d=2n \\ 
\frac{-1}{4\pi \left( 2n+1\right) r_{h}^{2n}R_{0}^{\frac{2n-1}{2}}}\left[ 
\frac{8Q^{\frac{2n+1}{2}}}{2n-1}\left( \frac{\left( n-1\right) \left(
2n-1\right) }{2}\right) ^{\frac{2n+1}{4}}+\left[ R_{0}r_{h}^{2}-2\left(
2n+1\right) \left( n-1\right) \right] r_{h}^{2n-1}R_{0}^{\frac{2n-1}{2}}%
\right] , & d=2n+1%
\end{array}%
\right. .
\end{equation}%
The specific heat also fallows as%
\begin{equation}
C_{Q}=\left\{ 
\begin{array}{ll}
\frac{\pi ^{\frac{2n-1}{2}}r_{h}^{2\left( n-1\right) }nR_{0}^{n-1}\left(
n-1\right) \left( 2n-1\right) \Psi _{1}}{2\Gamma \left( n+\frac{1}{2}\right)
\Phi _{1}}, & d=2n \\ 
\frac{\pi ^{n}r_{h}^{2n-1}R_{0}^{\frac{2n-1}{2}}n\left( 4n^{2}-1\right) \Psi
_{2}}{4\Gamma \left( n+1\right) \Phi _{2}}, & d=2n+1%
\end{array}%
\right.
\end{equation}%
in which we have used the following abbreviations 
\begin{eqnarray}
\Psi _{1} &=&4Q^{n}\left( \frac{\left( 2n-3\right) \left( n-1\right) }{2}%
\right) ^{\frac{n}{2}}+\left( 6n-4n^{2}+R_{0}r_{h}^{2}\right) r_{h}^{2\left(
n-1\right) }R_{0}^{n-1} \\
\Phi _{1} &=&-4Q^{n}\left( \frac{\left( 2n-3\right) \left( n-1\right) }{2}%
\right) ^{\frac{n}{2}}\left( 2n-1\right) +\left(
-6n+4n^{2}+R_{0}r_{h}^{2}\right) r_{h}^{2\left( n-1\right) }R_{0}^{n-1} 
\notag \\
\Psi _{2} &=&4Q^{\frac{2n+1}{2}}\left( \frac{\left( n-1\right) \left(
2n-1\right) }{2}\right) ^{\frac{2n+1}{4}}+\frac{\left( 2n-1\right) }{2}\left[
R_{0}r_{h}^{2}-2\left( 2n+1\right) \left( n-1\right) \right]
r_{h}^{2n-1}R_{0}^{\frac{2n-1}{2}}  \notag \\
\Phi _{2} &=&-8nQ^{\frac{2n+1}{2}}\left( \frac{\left( n+1\right) \left(
2n-1\right) }{2}\right) ^{\frac{2n+1}{4}}+\frac{\left( 2n-1\right) }{2}\left[
R_{0}r_{h}^{2}+2\left( 2n+1\right) \left( n-1\right) \right]
r_{h}^{2n-1}R_{0}^{\frac{2n-1}{2}}.  \notag
\end{eqnarray}%
We notice that in odd dimensions from $f\left( R_{0}\right) =R_{0}^{\frac{d}{%
2}\text{ }},$ $R_{0}$ can not get negative values for $d=$odd integer. The
detail can be seen in Appendix.

\subsubsection{The First Law of Thermodynamics}

As it was shown in Ref. \cite{3} the first law of thermodynamics in $f(R)$
gravity can be expressed as%
\begin{equation}
TdS-dE=PdV
\end{equation}%
in which $E$ is the Misner-Sharp \cite{13} energy stored inside the horizon
such that%
\begin{equation}
dE=\frac{1}{2\kappa }\left[ \frac{\left( d-2\right) \left( d-3\right) }{%
r_{h}^{2}}f_{R}+\left( f-Rf_{R}\right) \right] \mathcal{A}_{h}dr_{h},
\end{equation}%
$T=\frac{A^{\prime }}{4\pi }$ is the Hawking temperature, $S=\frac{2\pi 
\mathcal{A}_{h}}{\kappa }f_{R},$ is the entropy of the black hole $%
P=T_{r}^{r}=T_{0}^{0}$ is the radial pressure of matter fields at the
horizon and $dV=\mathcal{A}_{h}dr_{h}$ is the change of volume of the black
hole at the horizon. In the case of constant curvature i.e., $R=R_{0}$ one
gets%
\begin{equation}
dE=\frac{1}{2\kappa }\left[ \frac{\left( d-2\right) \left( d-3\right) }{%
r_{h}^{2}}\frac{d}{2R_{0}}+\left( 1-\frac{d}{2}\right) \right] R_{0}^{\frac{d%
}{2}}\mathcal{A}_{h}dr_{h}
\end{equation}%
which implies%
\begin{equation}
E=\frac{\left( d-2\right) }{4\kappa }\left[ \frac{d\left( d-3\right) }{%
r_{h}\left( d+1\right) R_{0}}-\frac{r_{h}}{d-1}\right] R_{0}^{\frac{d}{2}}%
\mathcal{A}_{h}.
\end{equation}%
Here we show that the first law of thermodynamics for the metric function
(25) is satisfied. Herein $P=-\frac{1}{4\pi }\left( \frac{\left( d-2\right)
\left( d-3\right) Q^{2}}{4r_{h}^{4}}\right) ^{\frac{d}{4}}$ and therefore
the right hand side reads%
\begin{equation}
PdV=-\frac{1}{4\pi }\left( \frac{\left( d-2\right) \left( d-3\right) Q^{2}}{%
4r_{h}^{4}}\right) ^{\frac{d}{4}}\mathcal{A}_{h}dr_{h}.
\end{equation}%
On the other side we have 
\begin{equation}
TdS-dE=A^{\prime }\frac{\mathcal{A}_{h}}{4\kappa }\frac{d\left( d-2\right) }{%
r_{h}}R_{0}^{\frac{d}{2}-1}dr_{h}-\frac{1}{2\kappa }\left[ \frac{\left(
d-2\right) \left( d-3\right) }{r_{h}^{2}}\frac{d}{2R_{0}}+\left( 1-\frac{d}{2%
}\right) \right] R_{0}^{\frac{d}{2}}\mathcal{A}_{h}dr_{h}.
\end{equation}%
We combine the latter with (46) and (50) to rewrite the first law as%
\begin{equation}
A^{\prime }\frac{1}{4\kappa }\frac{d\left( d-2\right) }{r_{h}}R_{0}^{\frac{d%
}{2}-1}-\frac{1}{2\kappa }\left[ \frac{\left( d-2\right) \left( d-3\right) }{%
r_{h}^{2}}\frac{d}{2R_{0}}+\left( 1-\frac{d}{2}\right) \right] R_{0}^{\frac{d%
}{2}}\mathbf{=}-\frac{1}{4\pi }\left( \frac{\left( d-2\right) \left(
d-3\right) Q^{2}}{4r_{h}^{4}}\right) ^{\frac{d}{4}}
\end{equation}%
or equivalently%
\begin{equation}
A^{\prime }=\frac{\left( d-3\right) }{r_{h}}-\frac{r_{h}}{d}R_{0}-\frac{8}{%
d\left( d-2\right) r_{h}^{d-1}R_{0}^{\frac{d}{2}-1}}\left( \frac{\left(
d-2\right) d-3Q^{2}}{4}\right) ^{\frac{d}{4}},
\end{equation}%
which is the derivative of the metric function at $r=r_{h}.$ This shows that
the first law of thermodynamics by using the generalized form of the entropy
for the Misner-Sharp energy is satisfied. To conclude this section of
thermodynamics we must admit that we don't feel the necessity of addressing
the second law. This originates from the fact that we are entirely in the
static gauge so that the entropy change is assumed trivially satisfied i.e. $%
\Delta S=0.$

\section{Conclusion}

A relatively simpler class of solutions within $f\left( R\right) $ gravity
is the one in which the scalar curvature $R$ is a constant $R_{0}$ (both $%
R_{0}>0$ and $R_{0}<0$). We have concentrated on this particular class with
the supplementary condition of zero energy-momentum trace. The general
spherically symmetric spacetime minimally coupled with nonlinear Yang-Mills
(YM) field is presented in all dimensions $\left( d\geq 4\right) $. The YM
field can even be considered in the power-law form in which the YM
Lagrangian is expressed by $L\left( F\right) \sim \left( F^{a}.F^{a}\right)
^{\frac{d}{4}}$. Since exact solutions in $f\left( R\right) $ gravity with
external matter sources, are rare, such solutions must be interesting. The
equation of state for effective matter is considered in the form $%
P_{eff}=\omega \rho $, which is analyzed in Appendix A. The general forms of 
$\omega \left( r\right) $ given in (A21) determine $\omega $ within the
ranges of $-1<\omega <\frac{1}{d-1}$ and $0<\omega <\frac{1}{d-1}$
respectively. The fact that $\omega <-1$ doesn't occur eliminates the
possibility of ghost matter, leaving us with the YM source and the scalar
curvature $R_{0}$. In case that the YM field vanishes $\left( Q\rightarrow
0\right) $ the only source to remain is the effective cosmological constant $%
\Lambda _{eff}=\frac{\left( d-2\right) R_{0}}{2d}$, which arises naturally
in $f\left( R_{0}\right) $ gravity. Another interesting result to be drawn
from this study is that the effective pressure $P_{eff}$ changes sign before
/ after a critical distance. Thus, it is not possible to introduce a simple $%
\omega =$constant, so that the pressure preserves its sign in the presence
of a physical field (here YM) in the entire spacetime. From cosmological
considerations the interesting case is when the critical distance lies
outside the event horizon. This is depicted in the projective plot (Fig. 2)
of the heat capacity versus horizon and the charge. Finally it should be
added that although $f(R)=R^{d/2}$ gravities face viability problems in
experimental tests the occurrence of sources may render them acceptable in
this regard.

\textbf{Energy conditions}

When a matter field couples to any system, energy conditions must be
satisfied for physically acceptable solutions. We follow the steps as given
in \cite{14}.

\subsection{$R_{0}>0$}

\textit{Weak Energy Condition (WEC):}

\bigskip The WEC states that%
\begin{eqnarray}
\rho &\geq &0,  \TCItag{A1} \\
\rho +p_{i} &\geq &0.  \notag
\end{eqnarray}

In which $\rho $ is the energy density and \ $p_{i}$ are the principal
pressure components given by

\begin{eqnarray}
\rho &=&-\tilde{T}_{0}^{0}=\frac{R_{0}}{2\pi d}\left( \frac{F^{\frac{d}{4}}}{%
R_{0}^{\frac{d}{2}}}+\frac{\left( d-2\right) }{8}\right) ,  \TCItag{A2} \\
p_{i} &=&\tilde{T}_{i}^{i}=\frac{R_{0}}{2\pi d}\left( \frac{2}{\left(
d-2\right) }\frac{F^{\frac{d}{4}}}{R_{0}^{\frac{d}{2}}}-\frac{\left(
d-2\right) }{8}\right) ,\text{ \ \ \ \ \ \ \ \ \ \ \ \ \ \ \ \ \ \ \ \ \ }%
i=2,\cdots ,(d-1),  \notag \\
p_{1} &=&\tilde{T}_{1}^{1}=-\frac{R_{0}}{2\pi d}\left( \frac{F^{\frac{d}{4}}%
}{R_{0}^{\frac{d}{2}}}+\frac{\left( d-2\right) }{8}\right) .  \notag
\end{eqnarray}%
Both conditions are satisfied. So WEC is held.

\textit{Strong Energy Condition (SEC):}

This condition states that

\begin{eqnarray}
\rho +\underset{i=1}{\overset{d-1}{\tsum }}p_{i} &\geq &0,  \TCItag{A3} \\
\rho +p_{i} &\geq &0.  \notag
\end{eqnarray}%
The second condition is satisfied but first condition implies that

\begin{equation}
\rho +\underset{i=1}{\overset{d-1}{\tsum }}p_{i}=\frac{R_{0}}{2\pi d}\left( 2%
\frac{F^{\frac{d}{4}}}{R_{0}^{\frac{d}{2}}}-\frac{\left( d-2\right) ^{2}}{8}%
\right) \geq 0  \tag{A4}
\end{equation}%
or consequently 
\begin{equation}
\left( 2\left( \frac{F}{R_{0}^{2}}\right) ^{\frac{d}{4}}-\frac{\left(
d-2\right) ^{2}}{8}\right) \geq 0.  \tag{A5}
\end{equation}%
By a substitution from (11) for $F$ one finds that for $r<r_{c}$ the
condition is satisfied in which 
\begin{equation}
r_{c}=\sqrt[d]{\frac{16}{\left( d-2\right) ^{2}}}\sqrt[4]{\frac{\left(
d-2\right) \left( d-3\right) Q^{2}}{4R_{0}^{2}}}.  \tag{A6}
\end{equation}

\textit{Dominant Energy Condition (DEC):}

In accordance with DEC, the effective pressure must not be negative. This
amounts to

\bigskip 
\begin{equation}
P_{eff}=\frac{1}{d-1}\underset{i=1}{\overset{d-1}{\tsum }}T_{i}^{i}=\frac{1}{%
\left( d-1\right) }\frac{R_{0}}{2\pi d}\left( \frac{F^{\frac{d}{4}}}{R_{0}^{%
\frac{d}{2}}}-\frac{\left( d-2\right) \left( d-1\right) }{8}\right) \geq 0, 
\tag{A7}
\end{equation}%
which for $r<\tilde{r}_{c}$ it is fulfilled in which%
\begin{equation}
\tilde{r}_{c}=\sqrt[d]{\frac{8}{\left( d-2\right) \left( d-1\right) }}\sqrt[4%
]{\frac{\left( d-2\right) \left( d-3\right) Q^{2}}{4R_{0}^{2}}}.  \tag{A8}
\end{equation}

\bigskip \textit{Causality Condition (CC):}

In addition to the energy conditions one can impose the causality condition

\begin{equation}
0\leq \frac{P_{eff}}{\rho }=\frac{\left( F^{\frac{d}{4}}R_{0}^{\frac{-d}{2}}-%
\frac{\left( d-2\right) \left( d-1\right) }{8}\right) }{\left( d-1\right)
\left( F^{\frac{d}{4}}R_{0}^{\frac{-d}{2}}+\frac{\left( d-2\right) }{8}%
\right) }<1.  \tag{A9}
\end{equation}%
This is equivalent to%
\begin{equation}
F^{\frac{d}{4}}R_{0}^{\frac{-d}{2}}-\frac{\left( d-2\right) \left(
d-1\right) }{8}>0  \tag{A10}
\end{equation}%
which for $r<\tilde{r}_{c}$ is satisfied.

Finally we introduce $\omega =\frac{P_{eff}}{\rho },$ given by

\begin{equation}
\omega =\frac{\left( \left( \frac{F}{R_{0}^{2}}\right) ^{\frac{d}{4}}-\frac{%
\left( d-2\right) \left( d-1\right) }{8}\right) }{\left( d-1\right) \left(
\left( \frac{F}{R_{0}^{2}}\right) ^{\frac{d}{4}}+\frac{\left( d-2\right) }{8}%
\right) },  \tag{A11}
\end{equation}%
which is bounded as%
\begin{equation}
-1\leq \omega <\frac{1}{d-1}.  \tag{A12}
\end{equation}%
It is observed that 
\begin{equation}
\left\{ 
\begin{array}{ccc}
0\leq \omega <\frac{1}{d-1} & \text{if} & r<\tilde{r}_{c} \\ 
-1\leq \omega <0 & \text{if} & \tilde{r}_{c}<r%
\end{array}%
\right. .  \tag{A13}
\end{equation}

\subsection{$R_{0}<0$}

As one may see, presence of $R_{0}^{\frac{d}{2}}$ in the definition of $\rho 
$ and $p_{i}$ imposes that $d\neq 2n+1$ where for $n=2,3,4,...$. For $d=4n$
we get%
\begin{eqnarray}
\rho &=&-\tilde{T}_{0}^{0}=\frac{-\left\vert R_{0}\right\vert }{8\pi n}%
\left( \frac{F^{n}}{R_{0}^{2n}}+\frac{2n-1}{4}\right) ,  \notag \\
p_{i} &=&\tilde{T}_{i}^{i}=\frac{-\left\vert R_{0}\right\vert }{8\pi n}%
\left( \frac{1}{2n-1}\frac{F^{n}}{R_{0}^{2n}}-\frac{2n-1}{4}\right) , 
\TCItag{A14} \\
p_{1} &=&\tilde{T}_{1}^{1}=\frac{\left\vert R_{0}\right\vert }{8\pi n}\left( 
\frac{F^{n}}{R_{0}^{2n}}+\frac{2n-1}{4}\right) ,  \notag
\end{eqnarray}%
\textit{WEC: }These expressions reveal that the condition $\rho \geq 0$ and $%
\rho +p_{i}\geq 0$ are not satisfied. Similarly the SEC is also violated and
since the source is exotic we shall not consider it any further here. A case
of interest for $R_{0}<0$ is the choice $d=4n+2$ for $n=1,2,3,...$ in which%
\begin{eqnarray}
\rho &=&-\tilde{T}_{0}^{0}=\frac{\left\vert R_{0}\right\vert }{4\pi \left(
2n+1\right) }\left( \frac{F^{\frac{2n+1}{2}}}{\left\vert R_{0}\right\vert
^{2n+1}}-\frac{n}{2}\right) ,  \TCItag{A15} \\
p_{i} &=&\tilde{T}_{i}^{i}=\frac{\left\vert R_{0}\right\vert }{4\pi \left(
2n+1\right) }\left( \frac{1}{2n}\frac{F^{\frac{2n+1}{2}}}{\left\vert
R_{0}\right\vert ^{2n+1}}+\frac{n}{2}\right) ,\text{ \ \ \ \ \ \ \ \ \ \ \ \
\ \ \ \ \ \ \ \ \ }i=2,\cdots ,(d-1),  \notag \\
p_{1} &=&\tilde{T}_{1}^{1}=-\frac{\left\vert R_{0}\right\vert }{4\pi \left(
2n+1\right) }\left( \frac{F^{\frac{2n+1}{2}}}{\left\vert R_{0}\right\vert
^{2n+1}}-\frac{n}{2}\right) .  \notag
\end{eqnarray}%
\textit{WEC: }$\rho \geq 0$ yields%
\begin{equation}
\frac{F^{\frac{2n+1}{2}}}{\left\vert R_{0}\right\vert ^{2n+1}}-\frac{n}{2}%
\geq 0  \tag{A16}
\end{equation}%
or%
\begin{equation}
r<\bar{r}_{c}  \tag{A17}
\end{equation}%
where%
\begin{equation}
\bar{r}_{c}=\sqrt[4n+2]{\frac{2}{n}}\sqrt[4]{\frac{n\left( 4n-1\right) Q^{2}%
}{\left\vert R_{0}\right\vert ^{2}}}.  \tag{A18}
\end{equation}

\textit{SEC: }The conditions are simply satisfied.

\textit{DEC: }This amounts to

\bigskip 
\begin{equation}
P_{eff}=\frac{1}{4n+1}\frac{\left\vert R_{0}\right\vert }{4\pi \left(
2n+1\right) }\left( \frac{F^{\frac{2n+1}{2}}}{\left\vert R_{0}\right\vert
^{2n+1}}+\frac{n}{2}+2n^{2}\right) \geq 0,  \tag{A19}
\end{equation}%
which is also satisfied.

\bigskip \textit{CC: }The causality condition implies

\begin{equation}
0\leq \frac{P_{eff}}{\rho }=\frac{\left( \frac{F^{\frac{2n+1}{2}}}{%
\left\vert R_{0}\right\vert ^{2n+1}}+\frac{n}{2}+2n^{2}\right) }{\left(
4n+1\right) \left( \frac{F^{\frac{2n+1}{2}}}{\left\vert R_{0}\right\vert
^{2n+1}}-\frac{n}{2}\right) }<1,  \tag{A20}
\end{equation}%
or equivalently 
\begin{equation}
\left\vert R_{0}\right\vert ^{2n+1}\frac{1+4n}{4}<F^{\frac{2n+1}{2}} 
\tag{A21}
\end{equation}%
which is satisfied for%
\begin{equation}
r<\breve{r}_{c}  \tag{A22}
\end{equation}%
where%
\begin{equation}
\breve{r}_{c}=\sqrt[4n+2]{\frac{4}{1+4n}}\sqrt[4]{\frac{n\left( 4n-1\right)
Q^{2}}{\left\vert R_{0}\right\vert ^{2}}}.  \tag{A23}
\end{equation}%
Here the state function $\omega =\frac{P_{eff}}{\rho }$ becomes

\begin{equation}
\omega =\frac{\left( \frac{F^{\frac{2n+1}{2}}}{\left\vert R_{0}\right\vert
^{2n+1}}+\frac{n}{2}+2n^{2}\right) }{\left( 4n+1\right) \left( \frac{F^{%
\frac{2n+1}{2}}}{\left\vert R_{0}\right\vert ^{2n+1}}-\frac{n}{2}\right) }, 
\tag{A24}
\end{equation}%
which is bounded as%
\begin{equation}
-1\leq \omega <\frac{1}{4n+1}.  \tag{A25}
\end{equation}%
One can show that%
\begin{equation}
\left\{ 
\begin{array}{ccc}
0\leq \omega <\frac{1}{4n+1} & \text{if} & r<\bar{r}_{c} \\ 
-1\leq \omega <0 & \text{if} & \bar{r}_{c}<r%
\end{array}%
\right. .  \tag{A26}
\end{equation}

\textbf{Figure Captions:}

\textbf{Fig. 1}: The plot of horizon radius $r_{h}$ in $4-$dimensions versus
mass $m$ for different charges, $Q=0$ (Fig. 1A) and $Q=1$ (Fig. 1C). We also
plot the heat capacity $C_{Q}$ versus the horizon radius for $Q=0$ (Fig. 1B)
and $Q=1$ (Fig. 1D). Fig. 1D, displays in particular the instability caused
by the nonzero charge.

\textbf{Fig. 2:} The 3-dimensional picture of $C_{Q}$ versus $r_{h}$ and $Q$
as projected into the ($r_{h},Q$) plane. The shaded region with $C_{Q}>0$
shows the thermodynamically stable region. From cosmological point of view
the region of interest is when the critical $r_{c}$ is outside the event
horizon. As shown, below the curve $r_{h}=r_{c}$ we obtain stability (dark)
regions. Above the curve $r_{h}=r_{c}$, the region is already inside the
black hole and no stability is expected.


\begin{thebibliography}{99}
\bibitem{1} T. P. Sotiriou and V. Faraoni, Rev. Mod. Phys. \textbf{82}, 451
(2010);

S. Nojiri and S. D. Odintsov, Physics Reports, \textbf{505, }59 (2011)..

\bibitem{2} S. H. Hendi, Phys. Lett. B \textbf{690}, 220 (2010).

\bibitem{3} S. H. Mazharimousavi and M. Halilsoy, Phys. Rev. D \textbf{84},
064032 (2011).

\bibitem{4} P. B. Yasskin, Phys. Rev. D \textbf{12}, 2212 (1975).

\bibitem{5} K. Bamba and S. D. Odintsov, Journal of Cosmology and
Astroparticle Physics \textbf{04}, 024 (2008);

K. Bamba, S. Nojiri, and S. D. Odintsov, Phys. Rev. D \textbf{77}, 123532
(2008).

\bibitem{6} T. Moon, Y. S. Myung and E. J. Son, Gen. Relativ. Gravit. 
\textbf{43}, 3079 (2011);

S. H. Mazharimousavi, M. Halilsoy and T. Tahamtan, Eur. Phys. J. C. \textbf{%
72}, 1851 (2012);

A. de la Cruz-Dombriz, A. Dobado and A. L. Maroto, Phys. Rev. D \textbf{80},
124011 (2009);

A. de la Cruz-Dombriz, A. Dobado and A. L. Maroto, Phys. Rev. D \textbf{83},
029903(E) (2011);

S. Nojiri and S. D. Odintsov, Phys. Lett. B \textbf{657}, 238 (2007);

G. J. Olmo and D. R.-Garcia, Phys. Rev. D \textbf{84}, 124059 (2011).

\bibitem{7} S. H. Mazharimousavi, M. Halilsoy and Z. Amirabi, Gen Relativ
Gravit, \textbf{42}, 261 (2010).

\bibitem{8} S. H. Mazharimousavi and M. Halilsoy, Phys. Lett. B \textbf{659}
471 (2008);

S. H. Mazharimousavi and M. Halilsoy, Phys. Lett. B \textbf{694,} 54 (2010).

\bibitem{9} M. Hassaine, C. Martinez, Class. Quantum Grav. \textbf{25,}
195023 (2008);

H. Maeda, M. Hassaine, C. Martinez, Phys. Rev. D \textbf{79,} 044012 (2009);

S. H. Hendi, H.R. Rastegar-Sedehi, Gen. Relativ. Gravit. \textbf{41,} 1355
(2009);

S. H. Hendi, Phys. Lett. B \textbf{677,} 123 (2009);

M. Hassaine and C. Mart\'{\i}nez, Phys. Rev. D \textbf{75,} 027502 (2007) .

\bibitem{10} V. Faraoni, Phys. Rev. D \textbf{83}, 124044 (2011).

\bibitem{11} A.\thinspace D. Dolgov and M. Kawasaki, Phys. Lett. B \textbf{%
573}, 1 (2003).

\bibitem{12} M. Akbar and R. G. Cai, Phys. Lett. B \textbf{635}, 7 (2006);

Y. Gong and A. Wang, Phys. Rev. Lett. \textbf{99}, 211301 (2007);

R. Brustein, D. Gorbonos and M. Hadad, Phys. Rev. D \textbf{79}, 044025
(2009);

G. Cognola, E. Elizalde, S. Nojiri, S. D. Odintsov and S. Zerbini, J.
Cosmol. Astropart. Phys. \textbf{0502}, 010 (2005);

I. Brevik, S. Nojiri, S. D. Odintsov and L. Vanzo, Phys. Rev. D \textbf{70},
043520 (2004).

\bibitem{13} C. W. Misner and D. H. Sharp, Phys. Rev. \textbf{136}, B571
(1964);

M. Akbar and R. G. Cai, Phys. Lett. B \textbf{648}, 243 (2007);

R. G. Cai, L. M. Cao, Y. P. Hu and N. Ohta, Phys. Rev. D \textbf{80}, 104016
(2009);

H. Maeda and M. Nozawa, Phys. Rev. D \textbf{77}, 064031 (2008);

M. Akbar and R. G. Cai, Phys. Rev. D \textbf{75}, 084003 (2007);

M. Akbar and R. G. Cai, Phys. Lett. B \textbf{635}, 7 (2006);

R. G. Cai, L. M. Cao and N. Ohta, Phys. Rev. D \textbf{81}, 084012 (2010).

\bibitem{14} S. W. Hawking and G. F. R. Ellis, The Large Scale Structure of
Space-Time, Cambridge University Press (1973);

M. Salgado, Class. Quant. Grav. \textbf{20}, 4551, (2003);

S. H. Mazharimousavi, O. Gurtug and M. Halilsoy, Int. J. Mod. Phys. D 
\textbf{18}, 2061 (2009).

\textbf{Appendix A}
\end{thebibliography}
\end{document}